\documentclass[aps,prd,twocolumn,
comment, 
superscriptaddress,groupedaddress,
amsmath,amssymb,floatfix,footinbib]
{revtex4}  
\usepackage{float}
\usepackage{graphicx}
\usepackage{bm}
\usepackage[usenames,dvipsnames]{color}
\usepackage{slashed}
\usepackage{hyperref}
\usepackage{slashed}
\usepackage{graphics}
\usepackage{multirow}
\usepackage{tablefootnote}
\usepackage{booktabs}
\usepackage{amsmath}
\usepackage[capitalise]{cleveref}
\usepackage{slashed}

\def\mpl{M_{\rm {Pl}}}

\begin{document}

\renewcommand{\(}{\left(}
\renewcommand{\)}{\right)}
\renewcommand{\{}{\left\lbrace}
\renewcommand{\}}{\right\rbrace}
\renewcommand{\[}{\left\lbrack}
\renewcommand{\]}{\right\rbrack}
\renewcommand{\Re}[1]{\mathrm{Re}\!\{#1\}}
\renewcommand{\Im}[1]{\mathrm{Im}\!\{#1\}}
\newcommand{\dd}[1][{}]{\mathrm{d}^{#1}\!\!\;}
\newcommand{\del}{\partial}
\newcommand{\nn}{\nonumber}
\newcommand{\ie}{i.e.\,}
\newcommand{\cf}{cf.\,}
\newcommand{\refeq}[1]{Eq.~(\ref{eq:#1})}
\newcommand{\refeqs}[2]{Eqs.~(\ref{eq:#1})-(\ref{eq:#2})}
\newcommand{\reffig}[1]{Fig.~\ref{fig:#1}}
\newcommand{\refsec}[1]{Section \ref{sec:#1}}
\newcommand{\reftab}[1]{Table \ref{tab:#1}}
\newcommand{\order}[1]{\mathcal{O}\({#1}\)}
\newcommand{\fv}[1]{\left(\begin{array}{c}#1\end{array}\right)}%

\def\tcb#1{\textcolor{blue}{#1}}
\def\tcr#1{\textcolor{red}{#1}}
\def\tcg#1{\textcolor{green}{#1}}
\def\tcc#1{\textcolor{cyan}{#1}}
\def\tcv#1{\textcolor{violet}{#1}}
\def\tcm#1{\textcolor{magenta}{#1}}
\def\tcpn#1{\textcolor{pink}{#1}}
\def\tcpr#1{\textcolor{purple}{#1}}
\definecolor{schrift}{RGB}{120,0,0}

\newcommand{\alphas}{\alpha_\mathrm{s}}
\newcommand{\alphae}{\alpha_\mathrm{e}}
\newcommand{\gfermi}{G_\mathrm{F}}
\newcommand{\GeV}{\,\mathrm{GeV}}
\newcommand{\MeV}{\,\mathrm{MeV}}
\newcommand{\amp}[1]{\mathcal{A}\left({#1}\right)}
\newcommand{\wilson}[2][{}]{\mathcal{C}_{#2}^{\mathrm{#1}}}
\newcommand{\bra}[1]{\left\langle{#1}\right\vert}
\newcommand{\ket}[1]{\left\vert{#1}\right\rangle}

\def\be{\begin{equation}}
\def\ee{\end{equation}}
\def\bea{\begin{eqnarray}}
\def\eea{\end{eqnarray}}
\def\bm{\begin{matrix}}
\def\em{\end{matrix}}
\def\bpm{\begin{pmatrix}}
    \def\epm{\end{pmatrix}}

{\newcommand{\lsim}{\mbox{\raisebox{-.6ex}{~$\stackrel{<}{\sim}$~}}}
{\newcommand{\gsim}{\mbox{\raisebox{-.6ex}{~$\stackrel{>}{\sim}$~}}}
\def\gev{{\rm \,Ge\kern-0.125em V}}
\def\tev{{\rm \,Te\kern-0.125em V}}
\def\mev{{\rm \,Me\kern-0.125em V}}
\def\ev{\,{\rm eV}}

\def\Mpl{M_{\rm Pl}}

\title{\boldmath  \color{schrift}{Stochastic gravitational wave from graviton bremsstrahlung in inflaton decay into massive spin 3/2 particles}}
\author{Diganta Das}
\author{Mihika Sanghi}
\author{Sourav}
\affiliation{Center for Computational Natural Sciences and Bioinformatics,
International Institute of Information Technology, Hyderabad 500 032, India}

\begin{abstract}
The detection of primordial gravitational waves would offer a direct evidence of inflation and valuable insights into the dynamics of the early universe. During post-inflation reheating period, when the inflaton coherently oscillates at the bottom of its potential, primordial stochastic gravitational waves may be sourced by its perturbative decay into particles of different spins. Assuming the behavior of the potential near the minimum as a polynomial $V(\phi)\sim \phi^k$, where $k\ge 2$, and treating the inflaton as coherently oscillating classical field, we calculate the decay of inflaton into a pair of spin $3/2$ particles accompanied by graviton emission. We numerically study the reheating dynamics and calculate the stochastic gravitational wave spectra. Our analysis shows that the gravitational wave spectra can offer insights into the microscopic physics during inflation. 
	
\end{abstract}

\maketitle
\section{Introduction}
Inflation, a quasi-exponential expansion phase of the early universe, was proposed partly to solve the magnetic monopole problem, the fine-tuning problem and the flatness problem \cite{Guth:1980zm, Starobinsky:1980te, Linde:1981mu, Albrecht:1982wi, Kazanas:1980tx, Sato:1980ac}. The expansion is attributed to a scalar field, called inflaton $\phi$. It occurs when the energy density of the universe is dominated by the potential of the inflaton, $V(\phi)$, and the field `slowly rolls' down towards the potential minimum. While the field approaches the potential minima, it coherently oscillates and dissipates its energy into relativistic particles. This leads to reheating of the universe, which sets the stage for  big bang nucleosynthesis.  

Other than the resolution of some of the key puzzles of the universe, what makes inflation so attractive is its elegant explanation of the universe's large-scale structure formation. It predicts that tiny quantum fluctuations that originate during the inflationary expansion are eventually stretched by inflation to observable scales \cite{Bardeen:1980kt, Mukhanov:1990me, Kiefer:2008ku, Sudarsky:2009za, Martin:2012ua,Mukhanov:1981xt, Mukhanov:1982nu, Hawking:1982cz, Guth:1982ec, Starobinsky:1982ee,Bardeen:1983qw, Mukhanov:1985rz}.  There are several observations that indirectly support the idea. For example, the nearly scale-invariant primordial power spectrum observed in the cosmic microwave background (CMB) by Planck and WMAP \cite{Planck:2018jri, Planck:2018vyg, WMAP:2003ivt}, as well as the presence of super-horizon correlations in the CMB anisotropies \cite{COBE:1992syq,Planck:2018vyg}, favor an early inflationary phase. Constraints on tensor-to-scalar ratio from BICEP/Keck observations \cite{BICEP:2021xfz} further restrict the possible inflationary scenarios. However, a direct probe of inflation is long-awaited and the detection of stochastic gravitational wave (GW) background originating during the inflation is a prime candidate.

GWs are the manifestations of the tensor fluctuations of the spacetime metric and can originate during or after inflation. Tensor perturbations can be sourced by quantum fluctuations during inflation \cite{Starobinsky:1979ty, Allen:1987bk, Sahni:1990tx, Turner:1993vb}, rapid particle production through parametric resonance during preheating \cite{Khlebnikov:1997di, Easther:2006gt, Dufaux:2007pt, Bethke:2013aba, Figueroa:2017vfa}, fluctuations in the thermal plasma during post-reheating radiation dominated universe \cite{Ghiglieri:2015nfa, McDonough:2020tqq, Ghiglieri:2020mhm, Ringwald:2020ist, Klose:2022knn, Klose:2022rxh, Ringwald:2022xif, Ghiglieri:2022rfp, Muia:2023wru, Drewes:2023oxg, Ghiglieri:2024ghm}, and primordial black hole evaporation \cite{Anantua:2008am, Gehrman:2022imk, Ireland:2023avg, Gehrman:2023esa, Choi:2024acs}. References \cite{Nakayama:2018ptw, Huang:2019lgd} proposed the production of GWs through graviton bremsstrahlung. Gravitons emerge as  quantum fluctuations about the homogeneous and isotropic background and couple universally with all fields involved in the decay processes. During the reheating era, perturbative decay of the inflaton is accompanied by graviton emissions which constitute the stochastic GW background. These high-frequency GWs can be detected by future gravitational wave detectors operating in the GHz band and may shed light on the inflationary microphysics. 

In the present paper we discuss graviton bremsstrahlung in the decay of inflaton into massive spin 3/2 particles. Spin 3/2 particles are one of the first dark matter candidates to be ever considered. Though the spin 3/2 particles appear naturally in local supersymmetry, or supergravity theories, as early as 1941 Rarita and Schwinger proposed  a Lagrangian of a relativistic free spin 3/2 field \cite{Rarita:1941mf} based on the earlier works by Fierz and Pauli \cite{Fierz:1939ix}. The spin 3/2, often known as Rarita-Schwinger (RS) field, is a vector-valued spinor field and is denoted as $\psi_\mu$ where $\mu$ is the Lorentz index and the spinor index is usually suppressed. Recently, the gravitational production of spin 3/2 particles has been discussed in Refs.~\cite{Kaneta:2023uwi, Casagrande:2023fjk,Kolb:2023ydq,Garcia:2020hyo}. In supergravity, inflaton decay into spin-$3/2$ particles can arise through effective interactions determined by the K\"ahler potential and supersymmetry-breaking sector~\cite{Endo:2006zj} \cite{Endo:2007sz}.

In this paper, we compute the reheating dynamics and stochastic GWs generated during inflaton decay into a pair of RS particles and gravitons. We consider two-body decay of $\phi$ to a pair of RS particles, and three-body decay of $\phi$ to a pair of RS particles and a graviton. For reheating and the GW production in inflaton decay to spin $0,1,1/2$ particles, see references \cite{Nakayama:2018ptw, Barman:2023ymn, Barman:2023rpg, Kanemura:2023pnv, Bernal:2023wus,Ghoshal:2022kqp, Tokareva:2023mrt,Choi:2024acs}. In the literature, gravitational bremsstrahlung decays are calculated in two approaches. In the first approach, assuming inflaton as a collection of particles at rest, the decay rates are calculated using standard QFT in the Minkowski space. Since the Hubble time is much longer compared to the time period of the oscillation of  inflaton, the cosmic expansion can be ignored. This assumption also allows for the adoption of linearized theory of gravity in which the spacetime metric is expanded about the flat Minkowski background. In the second approach, which we follow in this paper, the inflaton is assumed to be a homogeneous classical field which oscillates at the bottom of its potential \cite{Barman:2023rpg} \cite{Kanemura:2023pnv} \cite{Bernal:2023wus} \cite{Ichikawa:2008ne} \cite{Garcia:2020wiy} \cite{Jiang:2024akb}. Without resorting to a specific model, we assume the potential to be of polynomial type $V(\phi)\sim \phi^k$ for $k\ge 2$, a form that a large class of viable models behave near the bottom, assumed to be $\phi=0$. The exponent $k$ governs the reheating dynamics and can be constrained from the GW spectrum. We undertake a fully numerical computation of the reheating dynamics and obtain the GW spectrum that would appear today. We find that the characteristic spectra can provide insights into the inflationary potential as well as the parameters of the underlying Lagrangian. However, the signal remain below the sensitivity of the present and future detector sensitivities.

The paper is organized as follows: In section \ref{sec:formalism} we briefly explain the linearized theory of gravity and discuss the action describing the interaction between inflaton, graviton and the spin 3/2 field. After a review of the dynamics of inflation in \ref{sec:dynamics}, the decay processes are calculated in \ref{sec:decay} which is followed by a discussion on reheating dynamics in \ref{sec:reheating}. The GW spectrum originating from the decays is investigated in \ref{sec:GW}. We conclude with a summary in \ref{sec:summary}.

\section{Formalism \label{sec:formalism}}
To compute the interactions of graviton with the other fields, the spacetime is perturbed as
\be
g_{\mu\nu} \simeq \eta_{\mu\nu} + \frac{2}{\mpl} h_{\mu\nu}\, ,
\ee
where $\eta_{\mu\nu}$ is the Minkowski metric and $|h_{\mu\nu}|\ll 1$. Before quantization, the $h_{\mu\nu}$ are classical fields: they are functions of the spacetime coordinates $x^\mu = ({\bf x}, t)$ assigning a tensor at each point in spacetime, and transform as a rank-2 tensor under a background Lorentz transformation. Since they arise after splitting the metric into a fixed background and a perturbation, the diffeomorphism invariance of the non-linear Einstein-Hilbert action remains but is no longer manifest. The quadratic part of the action is however invariant under the linearized diffeomorphism. After gauge is fixed and additional conditions are imposed, the $h_{\mu\nu}$ have two propagating degrees of freedom corresponding to the two polarization degrees of freedom of gravitational waves. To quantize $h_{\mu\nu}$, they are promoted to operators, and their excitations are particles called the gravitons. 
The classical equations of motion of $h_{\mu\nu}$ follow from the action, which in our case is 
\be\label{eq:EH-2}
S = \int d^4 x \sqrt{-g} \bigg( \frac{\mpl^2}{2}R + \frac{1}{2}\partial^\mu\phi\partial_\mu\phi - V(\phi) + \mathcal{L}_{3/2} + \mathcal{L}_{\rm int} \bigg)\, .
\ee
Here $R$ is the Ricci scalar, $\Mpl$ is the reduced Planck mass, and the $g$ is the determinant of the metric tensor. The second and the third term of the integrand correspond to the kinetic term and the potential term of the inflaton. The fourth term represents the Lagrangian of a free massive RS field
\begin{equation}\label{eq:free-RS-field}
    \mathcal{L}_{3/2} = \bar{\psi}_\mu \big( i\gamma^{\mu\rho\nu} \partial_\rho + m_{3/2} \gamma^{\mu\nu} \big) \psi_{\nu}\, ,
\end{equation}
where $m_{3/2}$ is the mass of the RS particle, 
\be
\gamma^{\mu\nu} = \frac{1}{2} [\gamma^\mu, \gamma^\nu]\, ,
\ee
and 
\be
\gamma^{\mu\nu\rho} = \gamma^\mu\gamma^\nu\gamma^\rho - \eta^{\mu\rho}\gamma^\nu - \eta^{\nu\rho} \gamma^\mu + \eta^{\mu\nu} \gamma^\rho\, .
\ee
The Lagrangian \eqref{eq:free-RS-field} yields the equations of motion for spin 3/2 
\be
(i\slashed{\partial} - m_{3/2})\psi_\mu = 0\, ,\qquad \gamma^\mu\psi_\mu = 0\, ,\quad \partial^\mu\psi_\mu = 0\, .
\ee
The equations imply four physical polarization states $\pm 3/2$ and $\pm 1/2$.

The last term of the action in the  eq.~\eqref{eq:EH-2} represents the universal interaction of gravitons with inflaton and the $\psi_\mu$, as well as the couplings between the $\phi$ and the $\psi_\mu$ fields
\bea
\sqrt{-g} \mathcal{L}_{\rm int} &=& \frac{1}{\mpl} \big(h_{\mu\nu} T^{\mu\nu}_\phi +  h_{\mu\nu} T^{\mu\nu}_\psi \big) +\mathcal{L}_{\phi\psi\psi} \, .
\eea
Here the energy-momentum tensor of the $\phi$ and $\psi_\mu$, respectively, are 
\bea
T^{\mu\nu}_\phi &=& \partial^\mu\phi \partial^\nu\phi - g^{\mu\nu} \bigg( \frac{1}{2} \partial^\alpha\phi \partial_\alpha \phi - V(\phi) \bigg)\, ,\\
T^{\mu\nu}_{\psi} &=& -\frac{i}{4} \bar{\psi}_\rho \gamma^{(\mu} \overleftrightarrow{\partial}^\nu) \psi^\rho\, ,\nn\\ &+& \frac{i}{2} \bar{\psi}^{(\nu} \gamma^{\mu)} \overleftrightarrow{\partial}_\rho \psi^\rho + \frac{i}{2} \bar{\psi}^\rho \gamma^{(\mu} \overleftrightarrow{\partial}_\rho \psi^{\nu)}\, .
\eea

The Lagrangian $\mathcal{L}_{\phi\psi\psi}$ describes the interaction between the inflaton and the RS fields. Following \cite{Gherghetta:2024tob} we consider two types of interactions $\lambda_s\bar{\psi}_\mu \psi^\mu \phi$ and $\lambda_p\bar{\psi}_\mu \gamma_5\psi^\mu \phi$ where $\lambda_s$ and $\lambda_p$ are the coupling constants. If the RS field is massless, the non-derivative interactions are forbidden by the gauge symmetry of the massless RS field which removes unphysical spin 1/2 component. But as mentioned in the introduction, RS fields arise naturally in theories such as the $N=1$ supergravity, where they are known as gravitinos, superpartner of gravitons. After the super-Higgs mechanism, the gravitinos become massive by `eating' the spin 1/2 goldstinos. Once the mixing between the massive gravitinos and the goldstinos takes place, the gauge symmetry is broken and the non-derivative terms are allowed. However, to simplify our analysis, we remain agnostic to the underlying theory describing a massive RS field.

\section{Inflaton Dynamics \label{sec:dynamics}}
As mentioned in the introduction, the minimum of the potential is $\phi=0$ without any loss of generality, and a large class of models near the minimum assume the form
\begin{equation}\label{eq:poly_potential}
    V(\phi) = \lambda M_P^4 \left( \frac{\phi}{M_P} \right)^k, \quad \phi \ll M_P\, ,
\end{equation}
where $k\ge 2$. The constant $\lambda$ can be determined from the CMB data once the potential is specified. Since we are not concerned with the exact shape of the potential, we take $\lambda = 10^{-11}$ for our analysis. During the late stage of inflation or the early stage of reheating, the perturbative decay rate of $\phi$, $\Gamma_\phi\ll H$ so that the equation of motion of $\phi$ is
\be\label{eq:EoM-phi}
\ddot{\phi} + 3H\dot{\phi} + V^\prime(\phi) = 0\, ,
\ee
where $\dot{\phi}$ and $\ddot{\phi}$ are the first and second order time derivatives, respectively, and $V^\prime(\phi) = \partial V(\phi)/\partial\phi = k\lambda M_p^{4-k} \phi^{k-1}$. As the time period of the post-inflationary oscillation is sufficiently small compared to the Hubble time, the time dependence of $\phi$ can be split into two parts: an amplitude $\phi_0(t)$ that slowly decays due to cosmic expansion and energy dissipation, and a purely oscillatory part denoted by $\mathcal{P}(t)$
\begin{equation}\label{eq:factor-phi}
    \phi(t) = \phi_0(t) \mathcal{P}(t).
\end{equation}
The $\phi_0(t)$ is approximately constant during one oscillation period.
Substituting the decomposition in the equation of motion \eqref{eq:EoM-phi} gives a set of coupled equations for the slowly-varying amplitude $\phi_{0}(t)$, and for the rapidly oscillating component $\mathcal{P}(t)$:
\begin{eqnarray}
\ddot{\mathcal{P}} + k\lambda M_{p}^{4-k}\phi_{0}^{k-2}\mathcal{P}^{k-1} &=& 0\, ,\\
\dot{\phi}_{0} + 3H\dot{\phi}_{0} + k\lambda M_{p}^{4-k}\phi_{0}^{k-1}\langle\mathcal{P}^{k-1}\dot{\mathcal{P}}\rangle &=& 0\, .\
\end{eqnarray}
Here, $\langle \dots \rangle$ denote an average over a one oscillation period. Using the identity $\int_{0}^{1}\frac{dx}{\sqrt{1-x^{k}}}=\frac{\sqrt{\pi}\Gamma(\frac{1}{k})}{k\Gamma(\frac{1}{2}+\frac{1}{k})}$, the equation for $\mathcal{P}(t)$ can solved to obtain the time period
\begin{equation}
    T = \frac{4\sqrt{\pi}}{k\sqrt{2\lambda M_{p}^{4-k}\phi_{0}^{k-2}}} \frac{\Gamma(\frac{1}{k})}{\Gamma(\frac{1}{2}+\frac{1}{k})}\, .
\end{equation}
In terms of the effective mass of the inflaton, defined as
\be
m_{\phi}^{2}(t) \equiv \frac{\partial^2V(\phi)}{\partial\phi^2}\bigg|_{\phi_{0}}\, ,
\ee
the angular frequency can be written as
\begin{equation}
    \omega = \frac{2\pi}{T} = m_{\phi}\sqrt{\frac{\pi k}{2(k-1)}}\frac{\Gamma(\frac{1}{2}+\frac{1}{k})}{\Gamma(\frac{1}{k})}.
\end{equation}
Since $\mathcal{P}(t)$ is periodic, it can be written as linear superposition of sinusoidal functions as
\begin{equation}
    \mathcal{P}(t) = \sum_{n=-\infty}^{\infty}\mathcal{P}_{n}e^{-in\omega t}.
\end{equation}
The coefficients $\mathcal{P}_n$ are real so they satisfy the condition $\mathcal{P}_{-n} = \mathcal{P}_{n}^*$. Additionally,  $\langle\mathcal{P}\rangle=0$ requires that $\mathcal{P}_{0}=0$. If the potential is quadratic, {\it i.e.,} $k=2$, only the $n=\pm1$ modes exist. However, for any potential with $k>2$, higher-frequency modes ($|n|>1$) are generally non-zero. 

\section{Inflaton decay \label{sec:decay}}
We are primarily interested in two-body decay $\phi\to \psi_\mu\psi_\mu$ and three-body  bremsstrahlung decay $\phi \to \psi_\mu\psi_\mu h_{\mu\nu}$. To compute the decay rates, the following spinor sums are used
\bea
P_{ab} &=& \sum_{s=-3/2}^{+3/2} u_a(p,s) \bar{u}_b(p,s) = \left( \slashed{p} + m_{3/2} \right)\times \, \nn\\&&
\left(\eta_{ab} - \frac{1}{3} \gamma_a \gamma_b - \frac{2}{3} \frac{p_a p_b}{m^2_{3/2}}
+ \frac{p_a \gamma_b - p_b \gamma_a}{3m_{3/2}}
\right)\, \\
%
%
Q_{ab} &=& \sum_{s=-3/2}^{+3/2} v_a(p,s) \bar{v}_b(p,s) = \left( \slashed{p} - m_{3/2} \right)\, \nn\\ &&
\left(\eta_{ab} - \frac{1}{3} \gamma_a \gamma_b - \frac{2}{3} \frac{p_a p_b}{m^2_{3/2}}
- \frac{p_a \gamma_b - p_b \gamma_a}{3m_{3/2}}
\right)
\eea
The massless gravitons in the external states are represented by their polarization vectors. Gravitons have two polarization states represented by polarization tensors $\epsilon^1_{\mu\nu}$ and $\epsilon^2_{ \mu\nu}$ satisfying the following conditions: symmetric condition $\epsilon^{i\mu\nu} = \epsilon^{i\nu\mu}$, tracelessness $\eta^{\mu\nu} \epsilon^i_{\mu\nu}=0$, orthonormality $\epsilon^{i\mu\nu}\epsilon^{j\ast}_{\mu\nu}=\delta^{ij}$, and transversality $\omega_\mu \epsilon^{i\mu\nu}=0$, where $\omega_\mu = (E_\omega,{\bf \omega})$ is the the four-momentum of the graviton. The polarization sum relation reads
\be
\sum_{\rm pol} \epsilon^{\ast\mu\nu}\epsilon^{\rho\sigma} = \frac{1}{2}\bigg(\hat{\eta}^{\mu\rho}\hat{\eta}^{\nu\sigma}+\hat{\eta}^{\mu\sigma}\hat{\eta}^{\nu\rho} - \hat{\eta}^{\mu\nu}\hat{\eta}^{\rho\sigma} \bigg)\, ,
\ee
where
\be
\hat{\eta}^{\mu\nu} = \eta_{\mu\nu} - \frac{\omega_\mu \bar{\omega}_\nu + \omega_\nu \bar{\omega}_\mu }{\omega\cdot \bar{\omega}}\, , \quad \bar{\omega} = (E_\omega, -{\bf \omega})\, .
\ee

The Feynman diagrams of the decay processes are shown in \ref{3bodydiagrams}. The diagram on top left gives vanishing contribution since the graviton couples with derivative of the inflaton which has zero momentum. The diagram on the bottom right also vanishes by the tracelessness condition of the graviton. Deferring the details of the decay rate calculations to Appendix \ref{sec:decay-rates}, the results are presented here. Since the two operators considered, $\lambda_s\bar{\psi}_\mu \psi^\mu \phi$ and $\lambda_p\bar{\psi}_\mu \gamma_5\psi^\mu \phi$, do not interfere, we present the decay rates separately for each of them. When only $\bar{\psi}_\mu\psi^\mu \phi$ is present, the two-body $\phi\to\psi_\mu\psi_\mu$ decay rate is
\bea
\Gamma^{1\to2}_\phi &=& \frac{\lambda_s^2 m_{\phi} }{72 \pi} \Biggr[(k-1) (k+2) \Big(\frac{\omega}{m_{\phi}}\Big)^3 \sum_{n=1}^{\infty}|\mathcal{P}_n|^2  n^3 \,\nn\\   &\times& \bigg(\frac{ \sqrt{1 - 4 y_n^2} \left( 1 - 10 y_n^2 + 42 y_n^4 - 72 y_n^6 \right)}{ y_n^4 } \bigg)\Biggr]\, .\qquad
\eea
For $\bar{\psi}_\mu \gamma_5\psi^\mu \phi$ type operator, the two-body $\phi\to\psi_\mu\psi_\mu$ decay rate is
\bea
\Gamma^{1\to2}_\phi &=& \frac{\lambda_p^2 m_{\phi} }{72 \pi} \Biggr[(k-1) (k+2) \Big(\frac{\omega}{m_{\phi}}\Big)^3 \sum_{n=1}^{\infty}|\mathcal{P}_n|^2  n^3\, \nn\\ &\times& \bigg(\frac{ \sqrt{1 - 4 y_n^2} \left( 1 - 2y_n^2 + 10y_n^4 \right)}{ y_n^4 }\bigg)\Biggr]\, .
\eea
In both the expressions
\be
x_n = \frac{E_w}{n\omega}\, ,\qquad y_n = \frac{m_{3/2}}{n\omega}\, .
\ee
The infinite sum over $n$ in the decay rates means that the decay of the coherently oscillating inflaton field can be interpreted as the sum of decay of the infinite number of harmonic modes of the oscillating $\phi$. A harmonic labeled as $n$ have mass $n\omega$ and constitute a fraction $b_n$ of the total inflaton energy. To make this interpretation explicit, we write the rates as
\be
\Gamma^{1\to 2}_\phi = \sum_{n=1}^{\infty} b_n \Gamma_{\phi_n}^{1\to 2}\, ,
\ee
where $b_n$ are positive and they satisfy \footnote{We have numerically checked the positivity of $b_n$ as well as the relation \eqref{eq:bn-sum}. For analytical proof see \cite{Jiang:2024akb}.}
\be\label{eq:bn-sum}
\sum_{n=1}^{\infty }b_n = \sum_{n=1}^{\infty } (k+2)(k-1) \bigg( \frac{\omega}{m_\phi} \bigg)^2 n^2 |\mathcal{P}_n|^2 = 1\, .
\ee
\begin{figure}
    \begin{tabular}{cc}
        \includegraphics[width=0.4\linewidth]{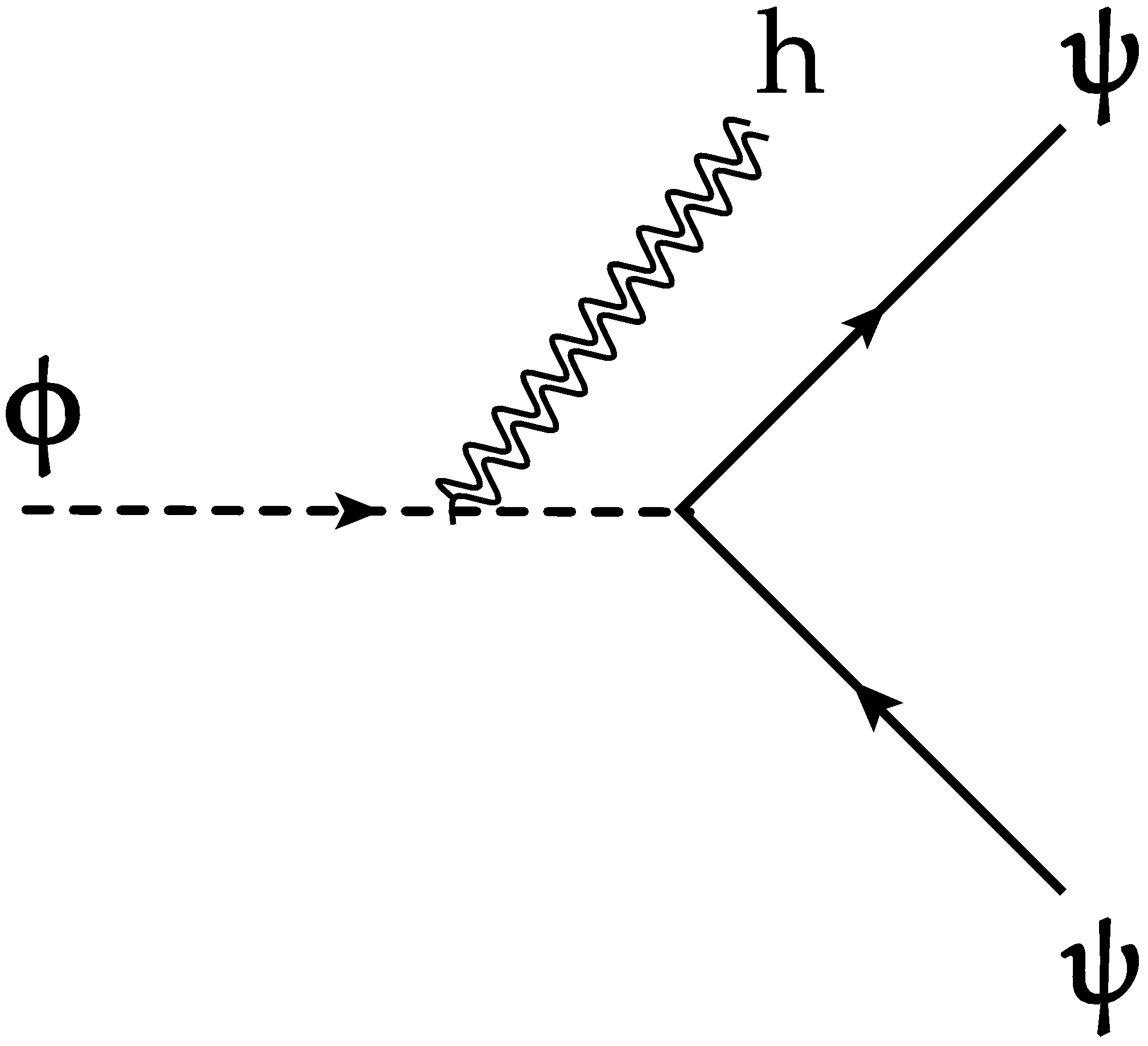} &
        \includegraphics[width=0.4\linewidth]{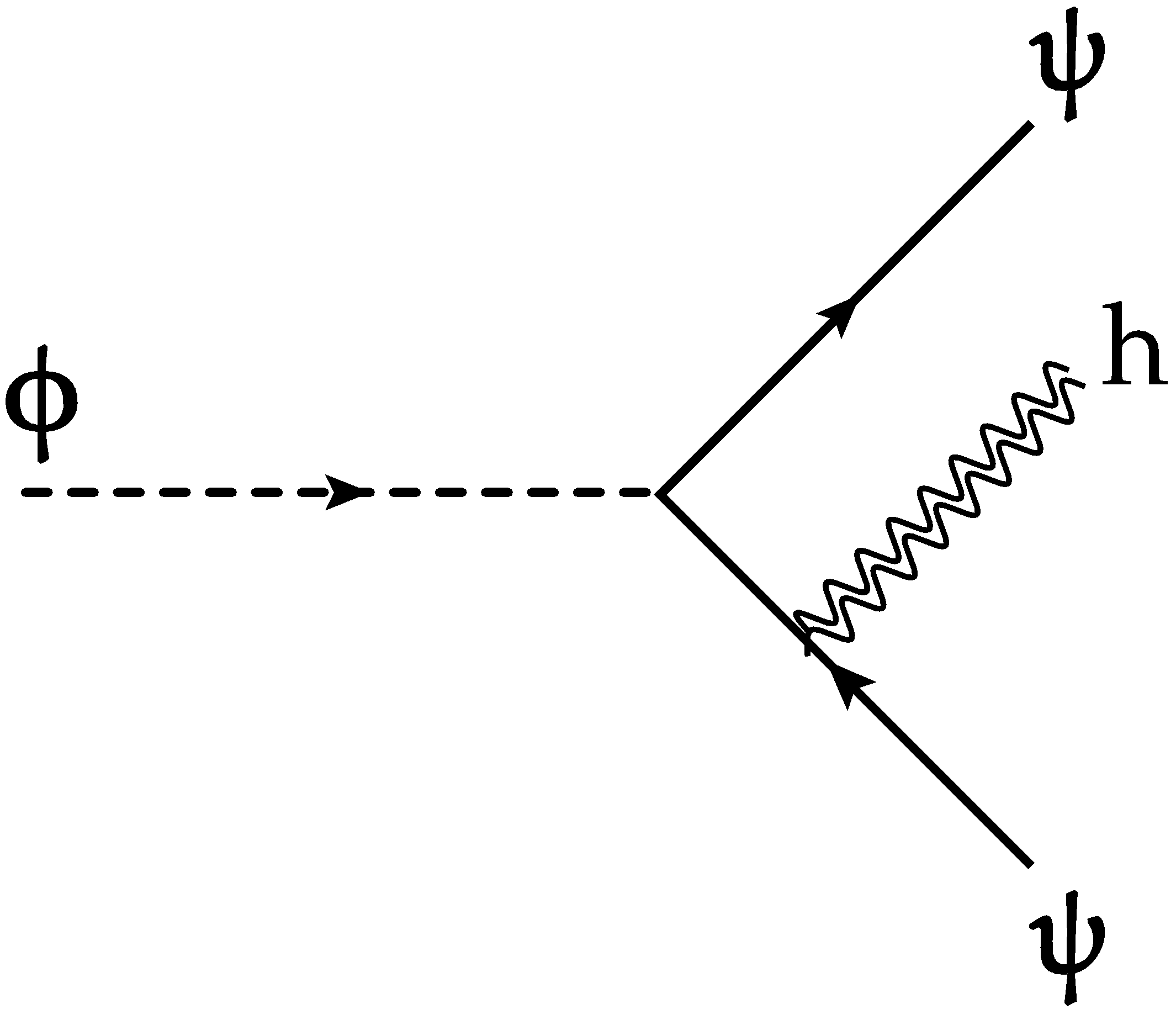} \\
        (a) & (b) \\[1em] 
        \includegraphics[width=0.4\linewidth]{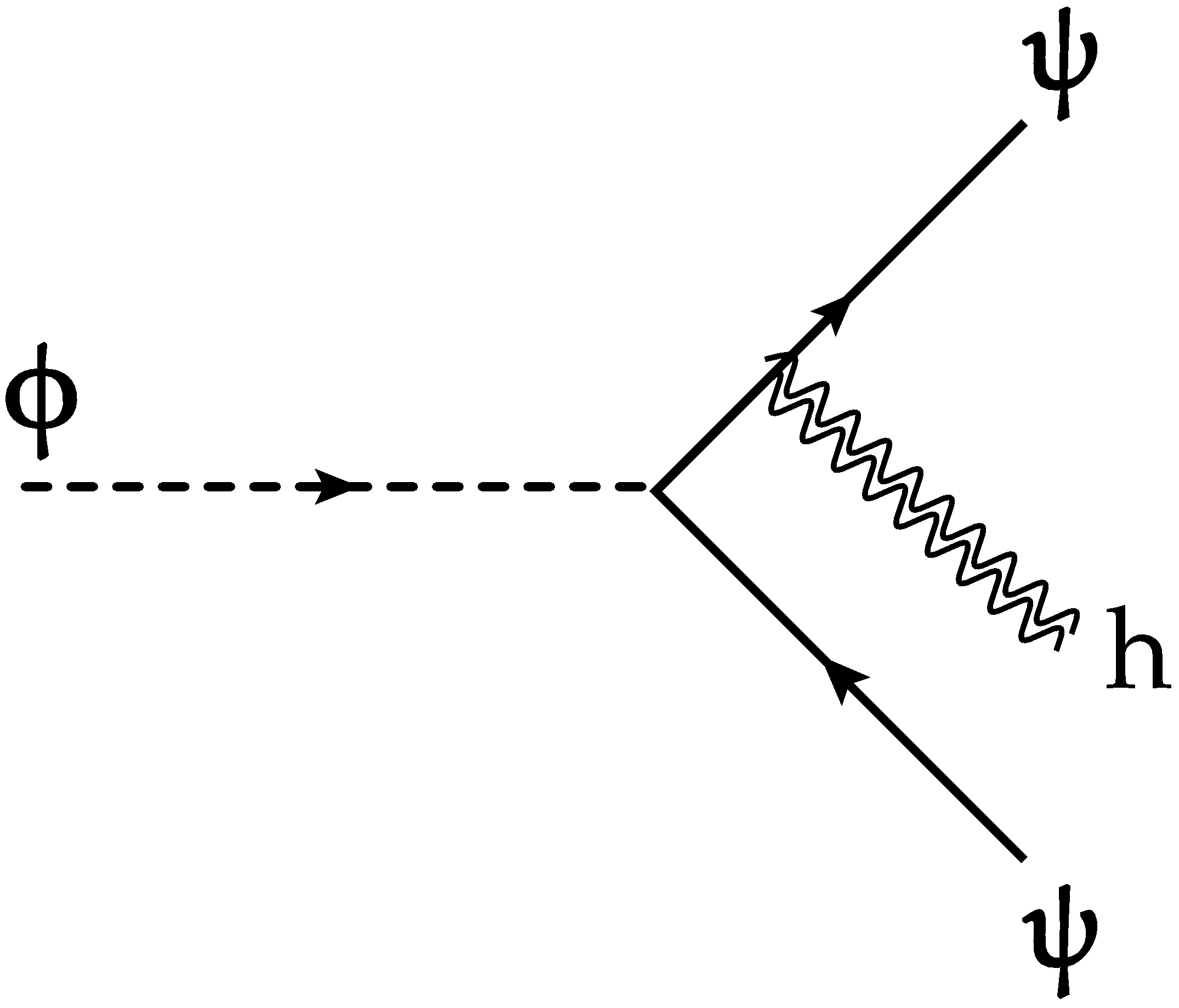} &
        \includegraphics[width=0.4\linewidth]{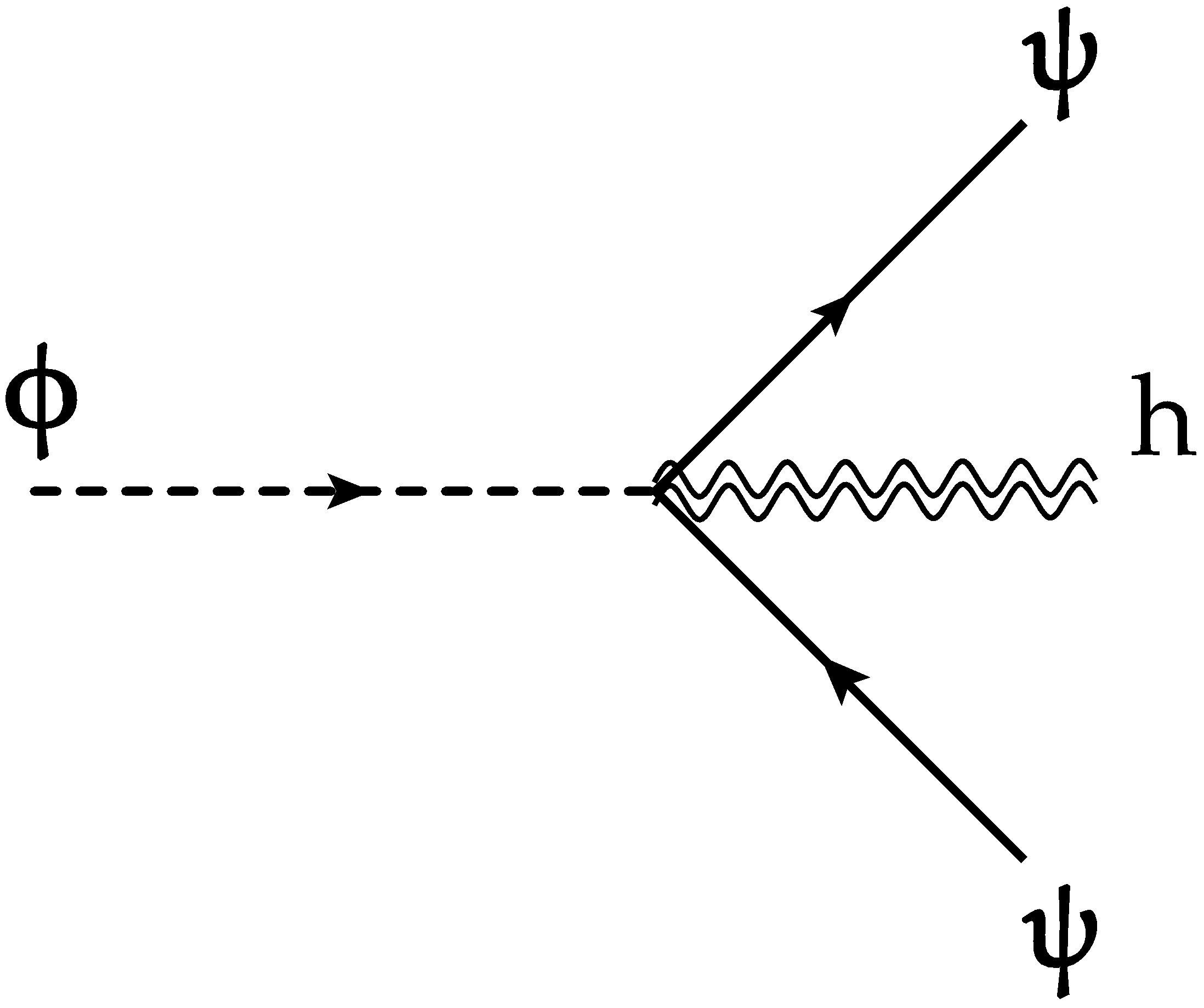} \\
        (c) & (d)
    \end{tabular}
    \caption{The Feynman diagrams for the three-body bremsstrahlung decay $\phi \to \psi_\mu{\psi_\mu} h_{\mu\nu}$. The $\psi_\mu$ is represented by $\psi$ and the graviton is represented by $h$. \label{eq:bn-sum}}
    \label{3bodydiagrams}
\end{figure}
Therefore, $b_n$ are normalized weight factors.

The three-body differential decay rates, $\phi\to \psi_\mu\psi_\mu h_{\mu\nu}$ for the $\bar{\psi}_\mu\psi^\mu \phi$ and $\bar{\psi}_\mu \gamma_5\psi^\mu \phi$ operators are
\begin{align}\label{eq:three-body-differential-rate}
\frac{d\Gamma^{1 \to 3}}{dE_\omega} &= \frac{\lambda_{s,p}^2}{2^{10}\cdot 3^5 \pi^3} \left(\frac{m_\phi}{M_p}\right)^2 (k+2)(k-1)\left(\frac{\omega}{m_\phi}\right)^4 \nonumber \\
&\times 
\Biggl[\sum_{n=1}^{\infty}  \frac{n^4 |\mathcal{P}_n|^2}{ x_n y_n^6 (\alpha^2 - 1)} \left( T_1^{s,p} + T_2^{s,p} + T_3^{s,p} + T_4^{s,p} \right) \Biggr]\, , 
\end{align}
subject to the kinematic constrain 
\be
0<x_n<\frac{1}{2}-2y_n^2 \, .
\ee
The terms $T_{1,2,3,4}^{s,p}$ depend on the types of operators considered, which are represented by the corresponding coupling constants $\lambda_s$ and $\lambda_p$. The expressions are presented in Appendix \ref{eq:aux-decay}. The interpretation of the infinite sum over $n$ is similar to the two body decay rate. Here also the differential rates can be written as
\begin{equation}
    \frac{d\Gamma^{1\to 3}}{dE_w} = \sum_{n=1}^\infty b_n \frac{d\Gamma^{1\to 3}_{\phi_n}}{dE_w}\, .
\end{equation}
%

\section{Reheating \label{sec:reheating}} 
Although the inflaton decay products considered here are spin-$3/2$ fermions, we follow Refs.~\cite{Nakayama:2018ptw, Barman:2023ymn, Barman:2023rpg, Kanemura:2023pnv, Bernal:2023wus, Tokareva:2023mrt} and adopt a phenomenological reheating description in which the energy transferred from inflaton decay is treated as contributing to an effective Standard Model radiation bath. The description allows one to meaningfully speak of a temperature $T$ and an effective number of relativistic degrees of freedom $g_*(T)$. At the level of effective field theory, a massive spin-$3/2$ field can admit higher-dimensional non-gravitational couplings to light degrees of freedom \cite{Garcia:2020hyo, Gherghetta:2024tob}, unless forbidden by an exact symmetry. Such couplings provide a plausible channel for energy transfer beyond purely gravitational interactions. A particularly favorable setup is non-gravitational coupling of the RS field with the Standard Model through a $d=5$ dimensional operator, $\mathcal{L}_{\rm int} \supset (c_d/\Lambda^{d-4})\mathcal{O}_d(\psi_\mu,\rm SM)$ where $c_d$ is the Wilson coefficient and $\Lambda$ is the ultraviolet cutoff scale. The resulting decay rate scales as $\Gamma_{3/2} \sim {c_5^2 m_{3/2}^3/\Lambda^2}$. Efficient thermalization requires the RS field to decay faster than the inflaton inverse-decay rate $\Gamma_{3/2} \gg \Gamma_\phi \sim \lambda_{s,p}^{2}\, m_\phi$. At the same time, consistency of the effective field theory demands $E/\Lambda \ll 1$, with the characteristic energy scale set by $E \sim m_\phi$. Both conditions are simultaneously satisfied in our benchmark scenario. For example, choosing $m_\phi = 10^{13}\,\mathrm{GeV}, m_{3/2} = 0.05\, m_\phi, \lambda_{s,p} = 10^{-6},\Lambda = 10^{15}\,\mathrm{GeV}, c_5 = 1$ one finds $\Gamma_{3/2} \gg \Gamma_\phi$ and $m_\phi/\Lambda\ll 1$. Therefore, efficient thermalization is possible while remaining well within the regime of validity of the effective field theory. We do not further attempt a microscopic description of the equilibration processes leading to this bath. Instead, we focus on the resulting gravitational wave signal within the standard reheating framework.

In this approach, the post-inflationary dynamics are described by coupled Boltzmann equations for the inflaton, radiation, and gravitational wave energy densities, providing a phenomenological mapping between inflaton decay and the subsequent radiation-dominated epoch. The results presented below should therefore be interpreted within this standard phenomenological reheating framework.

The evolutions of inflaton's energy density, $\rho_\phi$, and the SM radiation energy density, $\rho_R$, are given by a set of Boltzmann equations 
\begin{align}
\label{eq:Boltzmann-1}
& \dot{\rho}_\phi + 3H(1 + w_\phi)\rho_\phi = - (1 + w_\phi)\rho_\phi(\Gamma^{1\to2}_\phi + \Gamma^{1\to3}_\phi)\,,\\
\label{eq:Boltzmann-2}
& \dot{\rho}_R + 4H\rho_R = (1 + w_\phi)\rho_\phi \bigg( \Gamma^{1\to2}_\phi + \int \frac{d\Gamma^{1\to3}_\phi}{dE_w} \, dE_w\bigg)\, ,
\end{align}
where 
\be
\omega_\phi = \frac{k-2}{k+2}\, .
\ee
Note that the radiation density $\rho_R$ includes the GW energy density $\rho_{\rm gw}$. The Hubble parameter is given by the Friedmann equation
%
\be
\label{eq:Friedmann-1}
H^2 = \frac{1}{3\mpl^2} (\rho_\phi+\rho_R).
\ee
The set of equations \eqref{eq:Boltzmann-1}-\eqref{eq:Boltzmann-2} are solved fully numerically subject to the constrain \eqref{eq:Friedmann-1}. The initial conditions of simulation are determined by the values of the energy densities at the end of the inflation
\be\label{eq:rho-end}
\rho_{\rm end} = \rho_\phi(\phi_{\rm end})\, ,\quad\rho_R = 0\, ,\, ,
\ee
where $\phi_{\rm end}$ is the value of inflaton at the end of inflation. It is determined from the slow-roll conditions and depends on the choice of the potential. For the choices of potential functions \eqref{eq:poly_potential} it can be written in term of the parameter $k$ as
\be
\phi_{\rm end} = \sqrt{\frac{3}{8}}\Mpl \ln\bigg[ \frac{1}{2} + \frac{k}{3} (k+\sqrt{k^2+3}) \bigg]\, .
\ee
So the inflaton energy density at the end of inflation is
\be
\rho_{\rm end} = \rho(\phi_{\rm end}) = \frac{3}{2} \lambda \Mpl^4 \bigg( \frac{\phi_{\rm end}}{\Mpl} \bigg)^k\, .
\ee
The free parameters of our Lagrangian are the couplings $\lambda_{s,p}$ and the RS particle mass $m_{3/2}$. As the operators $\bar{\psi}_\mu\psi^\mu\phi$ and $\bar{\psi}_\mu\gamma\psi^\mu\phi$ do not interfere in the decay rates, for the rest of our analysis we consider only one of the couplings to be non-zero. 
\begin{figure}[h]
\includegraphics[width=0.4\textwidth]{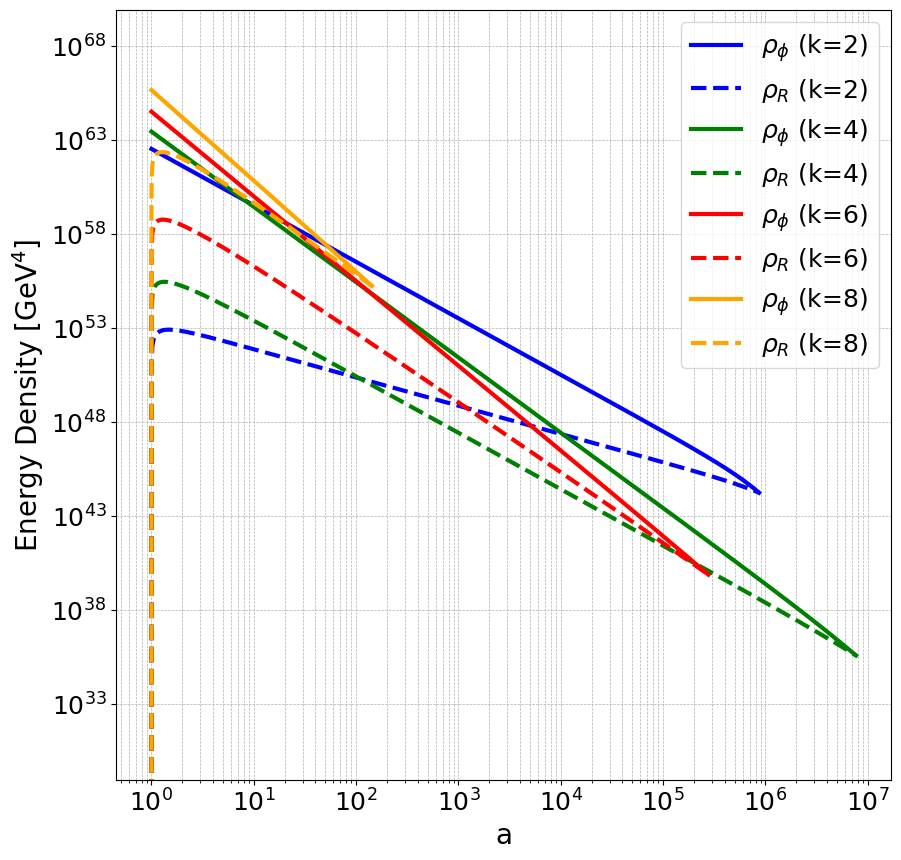}  
\caption{Evolution of inflaton energy density $\rho_\phi$ and radiation energy density $\rho_R$ as a function of the scale factor $a$ for $k = 2,4,6,8$. The coupling constant are set as follows: $\lambda_s=0$ and $\lambda_p=10^{-6}$ and $m_{3/2}=5\times 10^{-2}m_\phi$. A similar plot is obtained for non-zero $\lambda_s$ and $\lambda_p=0$.\label{fig:Reheating_plot_RS}}
\end{figure}

The evolutions of $\rho_\phi$ and $\rho_R$ as a function of the scale factor is shown in figure \ref{fig:Reheating_plot_RS} for $k=2,4,6,8$. The figure demonstrates that as the energy density $\rho_\phi$ is diluted due to dissipation of energy to the daughter field, the $\rho_R$ increases and reheats the universe. The intersection of $\rho_\phi$ and $\rho_R$ is defined as reheating. Depending on the shape of the potential $k$, reheating is achieved at different times. For the decays to take place $m_{3/2}<m_\phi/2$. In our analysis, the benchmark values of RS particles is restricted to $m_{3/2} \le 0.05 m_\phi$ ensuring that the spin-3/2 fermions are produced in the deep-relativistic to ultrarelativistic regime.

\section{Stochastic GW \label{sec:GW}}
The gravitons, after being produced in the three-body decay free streams across the universe without much interactions and constitute a statistically homogeneous and isotropic stochastic GW background. Due to the expansion of the universe, their frequency is redshifted and the energy density is diluted (scaling like radiation) by the time they reaches us. The observable of interest is the GW spectrum that describes the distribution of GW energy density $\rho_{\rm gw}$ across different frequency. It is defined with respect to the present day critical density of the universe $\rho_{\rm c,0}$
\be
\label{eq:omega-gw-0}
\Omega_{\rm gw,0} = \frac{1}{\rho_{\rm c,0}} \frac{d \rho_{\rm gw}(T_0) }{d\ln f} = \Omega_R^0 \frac{d \big( \rho_{\rm gw}(T_0)/\rho_{R}(T_0) \big) }{d\ln f}  \, ,
\ee
where $\Omega_R^0 h^2 = 2.47\times 10^{-5}$, where $h$ is the reduced Hubble constant, $T_0$ is the present day CMB temperature, $\rho_{gw}(T_0)$ and $\rho_R(T_0)$ are the present day energy densities, and $f$ is the present day frequency which related to $E_w$ as
\be
f=E_w \frac{a_{\rm rh}}{2\pi a_0}\, .
\ee
The $a_0$ is the current scale factor and $a_{\rm rh}$ is the scale factor at the time of reheating.  The factor $a_{\rm rh}/a_0$ accounts for the frequency redshift due to the expansion of the universe. Assuming no entropy production after the reheating until the present epoch, the comoving entropy conservation implies
\be\label{eq:a0-arh}
\frac{a_0}{a_{\rm rh}} = \frac{T_{\rm rh}}{T_0} \bigg( \frac{g_{\ast s}(T_{\rm rh})}{g_{\ast s}(T_0)} \bigg)^{1/3}\, ,
\ee
where $g_{\ast s}(T_0), g_{\ast s}(T_{\rm (rh)})$ are the relativistic entropy degrees of freedom. The reheating temperature $T_{\rm rh}$ is defined as $\rho_\phi(T_{\rm rh})=\rho_R(T_{\rm rh})$ and is determined from the numerical solutions of Boltzmann equations \eqref{eq:Boltzmann-1}-\eqref{eq:Boltzmann-2}. 

Since the time of reheating, the densities $\rho_{\rm gw}$ and $\rho_R$ scales as 
\be\label{eq:rho-gw-scaling}
\rho_{\rm gw} (T) = \rho_{\rm gw}(T_{\rm rh}) \bigg( \frac{a_{\rm rh}}{a} \bigg)^4\, ,
\ee
and
\be\label{eq:rho-r-scaling}
\rho_R(T) =   \frac{g_\ast(T)}{g_\ast(T_{\rm rh})}\bigg( \frac{T}{T_{\rm rh}} \bigg)^4 \rho_R(T_{\rm rh})\, ,
\ee
where $g_\ast(T_{\rm (rh)})$ is the number of relativistic degrees of freedom contributing to the energy density.  Using \eqref{eq:a0-arh}, \eqref{eq:rho-gw-scaling}, and \eqref{eq:rho-r-scaling} we can write
\bea\label{eq:rgw-rhoR-scaling}
\frac{\rho_{\rm gw}(T)}{\rho_{R}(T)} = \frac{\rho_{\rm gw}(T_{\rm rh})}{\rho_{R}(T_{\rm rh})} \frac{g_\ast(T_{\rm rh})}{g_\ast(T)} \bigg( \frac{g_{\ast s}(T)}{g_{\ast s}(T_{\rm rh})} \bigg)^{4/3}\, .
\eea
For numerical analysis we take $g_{\ast s}(T_{0}) = 3.91, g_{\ast }(T_0) = 3.36$,  $g_{\ast s}(T_{\rm rh}) = g_{\ast }(T_{\rm rh}) = 106.75$.
Using the equation \eqref{eq:rgw-rhoR-scaling} we find the present GW spectrum as
\bea
\Omega_{\rm gw, 0} = \Omega_{R,0} \frac{g_\ast(T_{\rm rh})}{g_\ast(T_0)} \bigg[ \frac{g_{\ast s}(T_0)}{g_{\ast s}(T_{\rm rh})} \bigg]^{\frac{4}{3}}\!\!\! \frac{1}{\rho_R(T_{\rm rh})} \frac{d \rho_{\rm gw}(T_{\rm rh})  }{d\ln E_w(T_{\rm rh})} \, .\quad
\eea
\begin{figure}[ht!]
\includegraphics[width=0.4\textwidth]{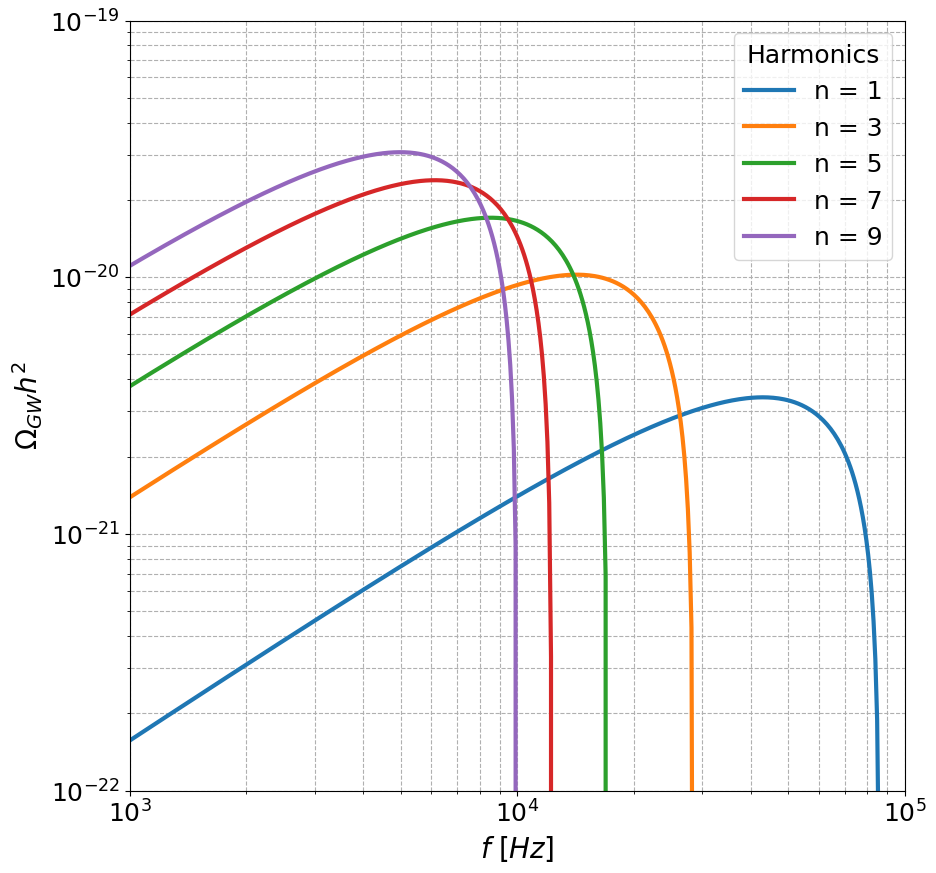}  
\caption{GW spectra coming from different harmonics for $k=6$. \label{fig:RS_GW_harmonics}}
\end{figure}
To calculate $d\rho_{\rm gw}/dE_w$ at the reheating temperature $T_{\rm rh}$ we numerically solve the equation
\begin{align}
\label{eq:Boltzmann-4}
&\frac{d}{dt} \left( \frac{d\rho_{\text{gw}}}{dE_w} \right) + 4H \frac{d\rho_{\text{gw}}}{dE_w} = (1+\omega_\phi) \frac{\rho_\phi}{m_\phi} \frac{d\Gamma^{1\to3}_\phi}{dE_w} \cdot E_w\, .\quad
\end{align}

\begin{figure}[ht!]
\includegraphics[width=0.4\textwidth]{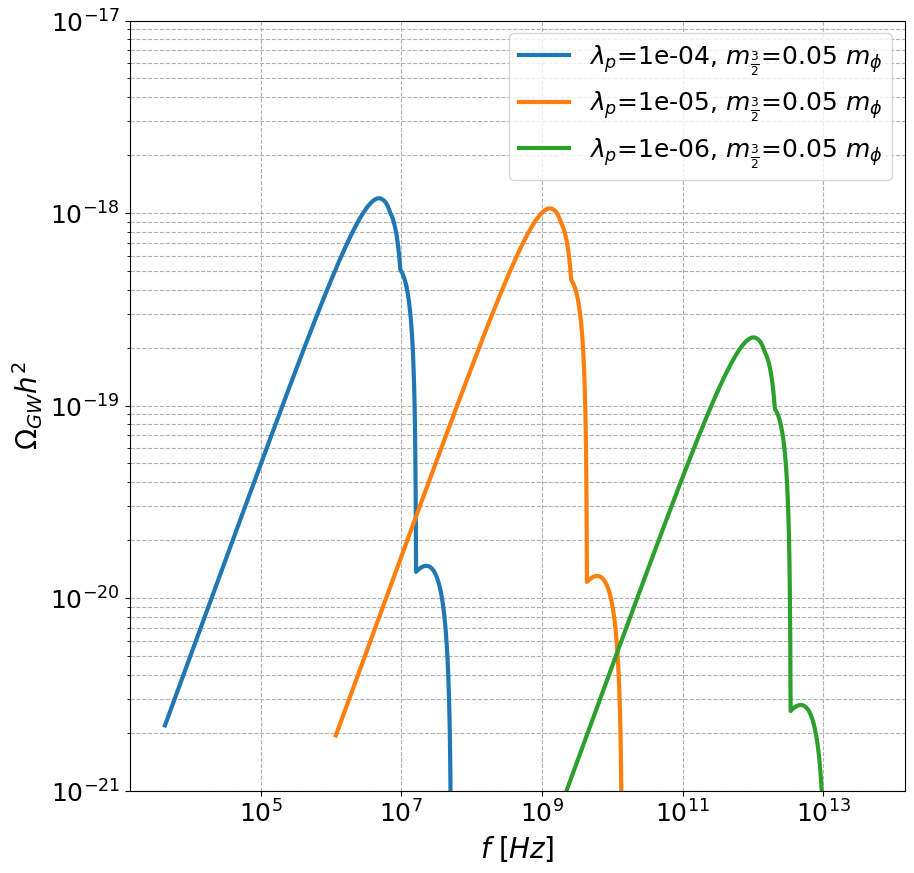}  
\caption{Sensitivity of the GW spectra for $k=6$ to the coupling constant $\lambda_p$ with $\lambda_s=0$.  \label{fig:variation_of_lp}}
\end{figure}

\begin{figure}[ht!]
\includegraphics[width=0.4\textwidth]{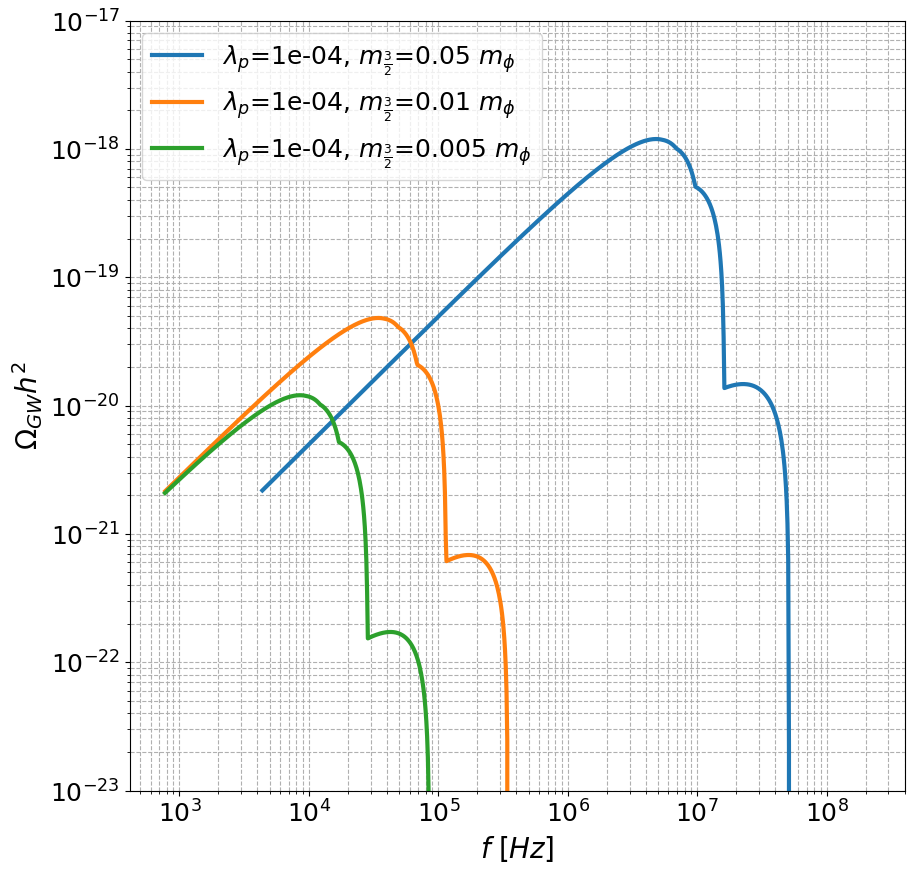} 
\caption{Sensitivity of the GW spectra for $k=6$ to masses of the RS field for a fixed $\lambda_p$ and $\lambda_s=0$.\label{fig:variation_of_mass}}
\end{figure}

Before we show the GW spectrum for different potentials, we present in figure \ref{fig:RS_GW_harmonics} the GW spectrum generated by the decay of the individual harmonics of the oscillating inflaton for $k=6$. In this figure, only the first nine harmonics are shown. We have observed that the final spectrum almost saturates with the first nine harmonics for the $k$ values studied.  The first thing to note from the figure is that for a given $k$, there is a hierarchy in the GW spectrum generated by the decay of the harmonics. The hierarchy is essentially is related to the hierarchy in decay rates of the harmonics. The feature was also noted in ref.~\cite{Jiang:2024akb} where graviton bremsstrahlung was discussed in inflaton decay into massless bosons and massless fermions. However, unlike in our case where the hierarchy is present in the harmonics for all $k$ values we have studied (apart from $k=2$ where only one harmonic contributes), in case of massless fermionic and bosonic reheating it is prominent only for $k=4$. The final spectrum for each $k$ are obtained by superimposing the individual harmonics and then taking the envelope. The final spectrum therefore shows the characteristic multi-peak features.

\begin{figure}[ht!]
\includegraphics[width=0.4\textwidth]{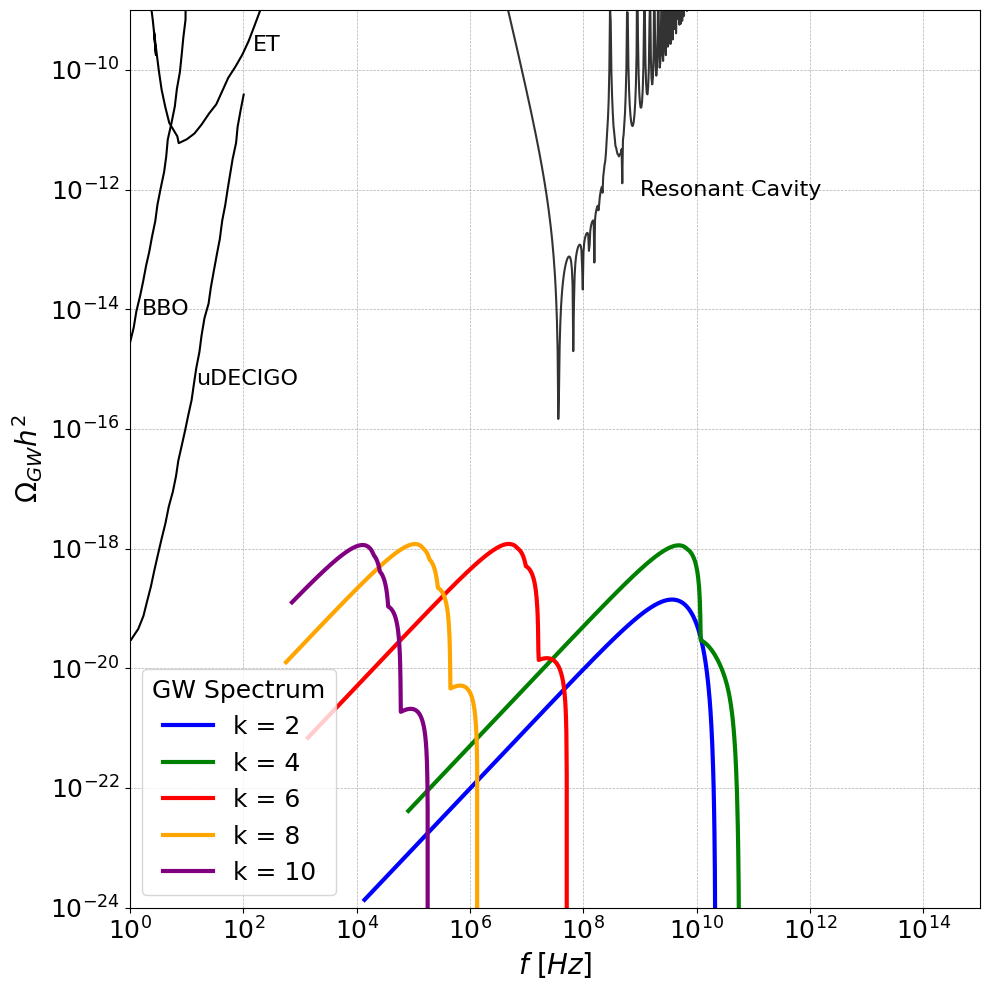} 
\includegraphics[width=0.4\textwidth]{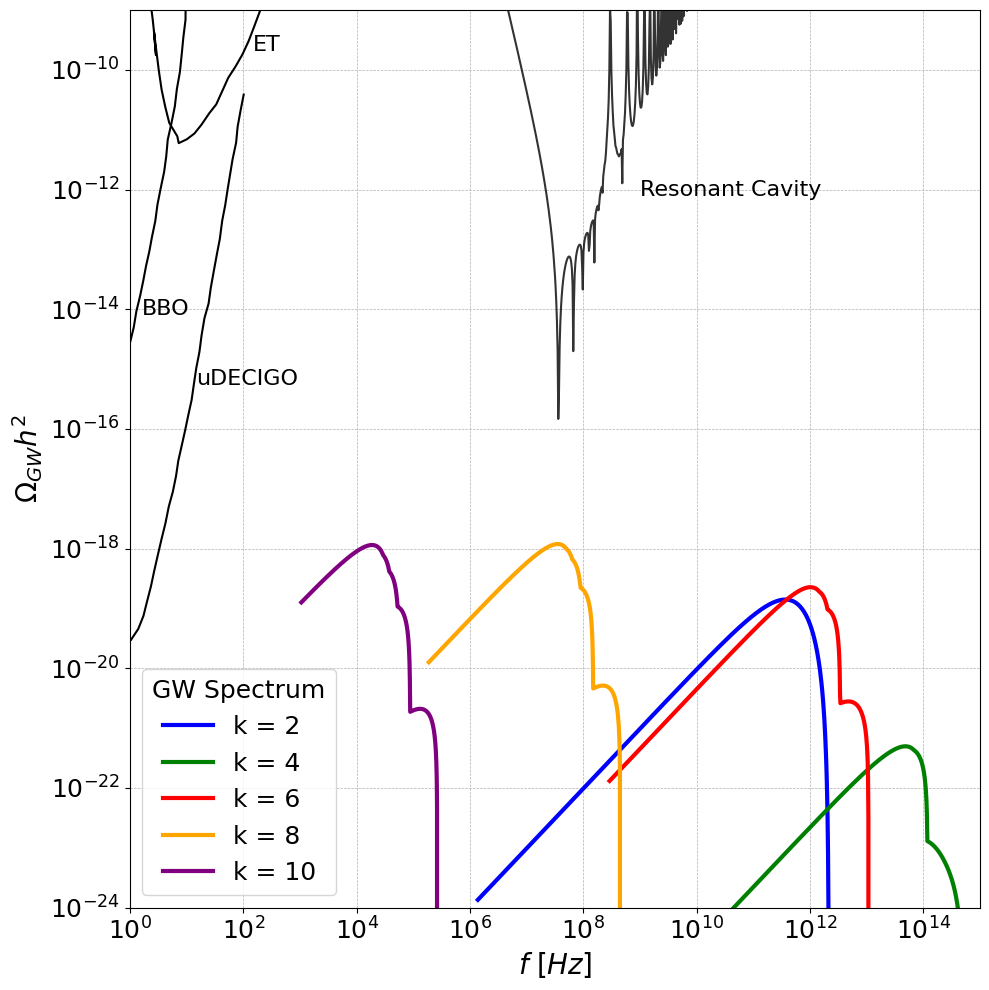}  
\caption{GW spectra for different potentials $k=2,4,6,8,10$ with two sets of parameters: $\lambda_p=10^{-4}, \lambda_s=0, m_{3/2}=0.05m_\phi$ (top) and $\lambda_p=10^{-6}, \lambda_s=0, m_{3/2}=0.05m_\phi$ (bottom). \label{fig:RS_GW_Plot_lp}}
\end{figure}

To show that the spectra are sensitive to the coupling constant $\lambda_{s,p}$, in figure \ref{fig:variation_of_lp} we present the spectrum for $k=6$ for different values of the coupling constant $\lambda_p$ and keeping $\lambda_s=0$. Decreasing the coupling constant $\lambda_p$ makes the spectrum peak at higher frequency. In figure \ref{fig:variation_of_mass} the sensitivity of the spectra to the mass $m_{3/2}$ is presented. All the figures shown are for $\lambda_s=0$ and non-zero $\lambda_p$. We have verified that almost similar figures are obtained with $\lambda_p=0$ and non-zero $\lambda_s$. Therefore, the gravitational wave spectra are not sensitive to the types of couplings between the inflaton and the RS field considered.

Finally, in figure \ref{fig:RS_GW_Plot_lp} we show the present day GW spectra for different $k$ for two sets of parameters: $\lambda_p=10^{-4}, \lambda_s=0, m_{3/2}=0.05m_\phi$ (top) and $\lambda_p=10^{-6}, \lambda_s=0, m_{3/2}=0.05m_\phi$ (bottom). The spectra are also compared with the sensitivity curves of some of the future GW detectors that include the ultimate Deci-Hertz Interferometer Gravitational Wave Observatory (uDECIGO) \cite{Seto:2001qf, Kudoh:2005as}, the Big Bang Observer (BBO) \cite{Crowder:2005nr, Corbin:2005ny}, the Resonance Cavities \cite{Berlin:2021txa, Herman:2022fau, Capdevilla:2024cby}, and the Einstein Telescope \cite{ET:2019dnz, Sathyaprakash:2012jk, Hild:2010id, Punturo:2010zz}. Clearly, the GW signals are out of reach of the future detectors. However, if detector sensitivity improves, the signals may be accessible and they can provide insights into the inflaton potential as well as the parameters of the RS Lagrangian.

\section{Summary \label{sec:summary}}
During the post-inflationary reheating of the universe, graviton emission in the perturbative decays of inflaton is unavoidable. As the gravitons propagate through the universe they produce gravitational waves of different amplitudes and frequencies. Their superpositions produce a stochastic background. The stochastic gravitational wave background is a fundamental observable in probing the inflationary microphysics. In this light, we have studied the gravitational waves produced during the perturbative decay of inflaton to a pair of spin 3/2 particles accompanied by graviton emission. Specifically, we have calculated the two body decay of inflaton into a pair of spin 3/2 particles,  and three body decay into a pair of spin 3/2 particles and a graviton. Solving the Boltzmann equations fully numerically, we have studied the evolution of the inflaton energy density and the energy density of the produced radiation. We have then obtained the gravitational wave spectrum for different potentials that would be present today. Though the spectrum is below the current and future observable limits, an observation may provide valuable information about the inflationary dynamics and reheating. 


\section*{Acknowledgment}
DD acknowledges the ANRF, Govt of India for support under the SRG grant (Sanction Order No. SRG/2023/001318) and the IIIT Hyderabad for the Seed Fund (No. IIIT/R\&D Office/Seed-Grant/2021-22/013). DD also acknowledges Guillem Domènech for kind hospitality during a stay at the Institut für Theoretische Physik, Leibniz Universität Hannover. SS and MS acknowledge Y.~Jiang and T.~Suyama for useful correspondence.

\section*{Data Availability}
The data that support the findings of this study are available from the corresponding author upon reasonable request.

\appendix


\section{Decay rate in classical field picture \label{sec:decay-rates}}
In this appendix we describe the decay of inflaton condensate to radiation following reference \cite{Garcia:2020wiy}. The evolution of the inflaton energy density is
\begin{equation}
\dot{\rho}_\phi + 3H(1 + w_\phi)\rho_\phi = -\Gamma_\phi(1 + w_\phi)\rho_\phi\, ,
\end{equation}
where the term in the right-hand side is the energy transfer, $\Delta E$, per spacetime volume, ${\rm Vol}_4$
\be\label{eq:source}
\Gamma_\phi(1 + w_\phi)\rho_\phi = \frac{\Delta E}{\rm Vol_4}\, .
\ee
The energy transfer is given by
\begin{equation}\label{eq:delta-E}
\Delta E \equiv \int \bigg( \Pi_f \frac{d^3\mathbf{p}_f}{(2\pi)^3 2p^0_f} \bigg) \bigg( \sum_f p^0_f \bigg) |\langle f | i \int d^4x \mathcal{L}_I | 0 \rangle|^2\, ,
\end{equation}
where the final state particles are denoted by $f$ and $p_f$ are the corresponding momenta, and $\mathcal{L}_I$ is the interaction Lagrangian. The modulus squared of the transition amplitude can be written as
\begin{equation}\label{eq:transition}
|\langle f|i \int d^4x \mathcal{L}_I |0 \rangle|^2 = \text{Vol}_4 \sum_{n=-\infty}^\infty |\mathcal{M}_n|^2 (2\pi)^4 \delta^4\bigg(p_n - \sum_f p_f\bigg),
\end{equation}
where $\mathcal{M}_n$ is the transition matrix element of the $n$-th mode and $p_n=(E_n,0)$, with $E_n$ as the energy of the $n$-th harmonic mode. To calculate the matrix element, the $\phi$ is treated as a time-dependent coefficient of the interaction. This allows us to factor out $\phi$ as equation \eqref{eq:factor-phi} and rewrite the matrix-element for the $n$-th harmonic as
\be\label{eq:MnMnp}
|\mathcal{M}_n|^2 = \phi_0^2|\mathcal{P}_n|^2 |\mathcal{M}_n^\prime|^2\, .
\ee
Using \eqref{eq:delta-E}, \eqref{eq:transition}, and \eqref{eq:MnMnp} we obtain the expression of decay rate from \eqref{eq:source} as
\bea\label{eq:master-formula}
\Gamma_\phi &=& \frac{1}{(1+\omega_\phi)\rho_\phi} \int \bigg( \Pi_f \frac{d^3\mathbf{p}_f}{(2\pi)^3 2p^0_f} \bigg) \bigg( \sum_f p^0_f \bigg) \,,\nn\\
&\times& \sum_{n=-\infty}^\infty \phi_0^2|\mathcal{P}_n|^2 |\mathcal{M}_n^\prime|^2 (2\pi)^4 \delta^4\bigg(p_n - \sum_f p_f\bigg).\quad\quad
\eea
The final expressions of two- and thee-body decay rates shown in the main texts follow from the equation \eqref{eq:master-formula}. In the final expressions we have made the replacements
\be
\omega_\phi = \frac{k-2}{k+2}\, ,\quad \frac{\rho_\phi}{\phi_0^2} = \frac{m_\phi^2}{k(k-1)}\, .
\ee

\section{Auxiliary expressions for decay rates \label{eq:aux-decay}}
The three-body differential rates are presented in equation \eqref{eq:three-body-differential-rate}. Here we present the expressions for $T^{s,p}_{1,2,3,4}$. The suffixes $s,p$ indicates the types of coupling considered. When only the $\lambda_s$ coupling between inflaton and RS fields is present, the decay rate is written in terms of $T_{1,2,3,4}^{s}$ 
\begin{widetext}
\bea
T_1^{s} &=& \alpha^5 (2x_n-1) \biggl[ -8(456x_n^2-500x_n+151)y^6 + 2592(1-2x_n)y_n^8 - 2(2x_n-1)(x_n(205x_n-172)+93)y_n^4 \nonumber \\
& \qquad &- 2(x_n(5x_n-7)+8)(y_n-2x_ny_n)^2 + x_n(2x_n-1)^3(3x_n-2) \biggr]\, \\
T_2^{s} &=& - 3\alpha \biggl[ (1-2x_n)^4 x_n^2 + 64(1398x_n-539)y_n^{10} - 32(x_n+1)(585x_n-287)y_n^8  - 20736y_n^{12} \, \nonumber\\ &\qquad& +16(2x_n-1)((x_n-186)x_n+28)y_n^6 
 + 2(1-2x_n)^2(x_n(527x_n-964)+115)y_n^4\, \nonumber\\ &+& 2x_n(2x_n-1)^3(37x_n-27)y_n^2 \biggr]\, ,\\
 T_3^s &=& 2\alpha^3 \biggl[ 31104(2x_n-1)y_n^{10} - 720(2x_n(22x_n-9)-3)y_n^8 - 4(2x_n-1)(2x_n(63x_n-50)+425)y_n^6 \nonumber \\
& \qquad & + 2(1-2x_n)^2(x_n(485x_n-884)+15)y_n^4  + 4(2x_n-1)^3(x_n(29x_n-22)+2)y_n^2 + (1-2x_n)^4x_n \biggr]\, , \\
T_4^s &=& - 12(\alpha^2-1)y_n^2 \biggl[ 24(125-374x_n)y_n^8 + 48(x_n(25x_n+14)-12)y_n^6 
 + 2592y_n^{10} - 16(2x_n-1)(x_n(62x_n-75)+8)y_n^4 \nonumber\, , \\
& \qquad & - 2(x_n(126x_n-223)+45)(y_n-2x_ny_n)^2  - (2x_n-1)^3(x_n(12x_n-17)+7) \biggr]  \log\left(\frac{1-\alpha}{1+\alpha}\right) \, ,
\eea
where
\be
\alpha = \sqrt{1-\frac{4 y_n^2}{1-2 x_n}}\, .
\ee
\end{widetext}

\begin{widetext}
When only the $\lambda_p$ coupling is present, the three-body decay rate is written in terms of $T_{1,2,3,4}^{p}$
\bea
T_1^p &=& \alpha^5 (2x_n-1) \biggl[ -2(2x_n-1)(376x_n^2 - 626x_n + 253)y_n^4 - 216(x_n(8x_n-9)+3)y_n^6 \nonumber \\
& \qquad& + 2(5x_n(4x_n+5)-19)(y_n-2x_ny_n)^2 + x_n(2x_n-1)^3(3x_n-2) \biggr]\, , \\
T_2^p &=& - 3\alpha \biggl[ (1-2x_n)^4 x_n^2 + 192(53-24x_n)y_n^{10} - 32(x_n(4939x_n-3515)+532)y_n^8 \nonumber \\
& \qquad & - 8(2x_n-1)(x_n(5122x_n-3423)+404)y_n^6 - 50(1-2x_n)^2(6(x_n-2)x_n+5)y_n^4 \nonumber \\
& \qquad & + 2(2x_n-1)^3(17x_n(4x_n+1)-25)y_n^2 + 11520y_n^{12} \biggr]\, , \\
T_3^p &=& 2\alpha^3 \biggl[ -144(526x_n^2-438x_n+83)y_n^8 - 12(2x_n-1)(2822x_n^2-2630x_n+565)y_n^6\, \nonumber \\
& \qquad & - 2(1-2x_n)^2(x_n(1501x_n-2057)+637)y_n^4 + 2(2x_n-1)^3(x_n(92x_n+13)-28)y_n^2 + (1-2x_n)^4x_n \biggr]\, , \\
T_4^p &=& - 12(\alpha^2-1)y_n^2 \biggl[ 24(14x_n-45)y_n^8 + 4(x_n(4385x_n-3237)+552)y_n^6 + 1440y_n^{10} \nonumber \\
& \qquad & + 4(2x_n-1)(x_n(1187x_n-881)+140)y_n^4 + 2(x_n(8x_n-35)+13)(y_n-2x_ny_n)^2 \nonumber \\
& \qquad & - x_n(2x_n-1)^3(20x_n-7) \biggr] \log\left(\frac{1-\alpha}{1+\alpha}\right)\, .
\eea
\end{widetext}

\end{document}